\documentclass[11pt]{article}
\usepackage{amsfonts,amssymb,amsmath,amsthm}
\usepackage{fullpage}
\usepackage{mathptmx}

\usepackage{todonotes}

\usepackage[colorlinks=true,citecolor=blue,urlcolor=blue,linkcolor=blue,bookmarksopen=true]{hyperref}

\newcommand{\F}{\mathbb{F}}

\newcommand{\eps}{\varepsilon}

\title{{\bf An improved bound on the fraction of correctable deletions}\thanks{A preliminary conference version of this paper~\cite{BG-soda16}, with a weaker bound of $1-2/(k+1)$ on fraction of correctable deletions, was presented at the 2016 ACM-SIAM Symposium on Discrete Algorithms (SODA) in January 2016.}}
\date{February 2016}

\author{Boris Bukh\thanks{Department of Mathematical Sciences, Carnegie Mellon University, Pittsburgh, PA 15213, USA. Email: {\tt bbukh@math.cmu.edu.} Supported in part by U.S. taxpayers through NSF grant DMS-1201380.} \and Venkatesan Guruswami\thanks{Computer Science Department, Carnegie Mellon University, Pittsburgh, PA 15213. Email: {\tt guruswami@cmu.edu.} Supported in part by NSF grants CCF-1422045 and CCF-0963975.} \and Johan H{\aa}stad\thanks{School of Computer Science and Communication, Royal Institute of Technology, Stockholm, Sweden. Email: {\tt johanh@kth.se}. Supported in part by Swedish Research Council.  Work done while visiting the Simons Institute in Berkeley.}}

\makeatletter
\def\TODO{\@ifnextchar[{\TODO@with}{\marginnote{TODO}}}
\def\TODO@with[#1]{\marginnote{#1}}
\makeatother

\newtheorem{theorem}{Theorem}
\newtheorem{lemma}[theorem]{Lemma}

\theoremstyle{definition}
\newtheorem{remark}{Remark}

\newcommand*{\abs}[1]{\lvert #1\rvert}                           
\newcommand*{\eqdef}{\stackrel{\text{\tiny{def}}}{=}}            
\newcommand*{\veps}{\varepsilon}                                 
\DeclareMathOperator{\wspan}{span}                               
\DeclareMathOperator{\len}{len}                                  
\DeclareMathOperator{\LCS}{LCS}                                  
\newcommand*{\0}{\textsf{0}}
\newcommand*{\1}{\textsf{1}}
\newcommand*{\2}{\textsf{2}}
\newcommand*{\3}{\textsf{3}}

\newcommand\drawsymbol[3]{\node at (#1,#2) {\tiny #3};}

\def\xspacing{0.3}
\def\xmetaspacing{3}

\newcommand\draweightsymbols[3]{
  \def\tempx{#1}
  \def\tempy{#2}
  \draweightsymbolscontinued#3
}

\def\draweightsymbolscontinued#1,#2,#3,#4,#5,#6,#7,#8{
    \drawsymbol{\tempx+\xspacing*0}{\tempy}{#1}
    \drawsymbol{\tempx+\xspacing*1}{\tempy}{#2}
    \drawsymbol{\tempx+\xspacing*2}{\tempy}{#3}
    \drawsymbol{\tempx+\xspacing*3}{\tempy}{#4}
    \drawsymbol{\tempx+\xspacing*4}{\tempy}{#5}
    \drawsymbol{\tempx+\xspacing*5}{\tempy}{#6}
    \drawsymbol{\tempx+\xspacing*6}{\tempy}{#7}
    \drawsymbol{\tempx+\xspacing*7}{\tempy}{#8}
}

\newcommand\drawmetasymbol[3]{
  \ifx1#3
   \draweightsymbols{#1*\xmetaspacing}{#2}{0,0,0,0,1,1,1,1}
  \fi
  \ifx2#3
    \draweightsymbols{#1*\xmetaspacing}{#2}{0,0,1,1,0,1,1,1}
  \fi
  \ifx3#3
    \draweightsymbols{#1*\xmetaspacing}{#2}{0,1,0,1,0,1,0,1}
  \fi
  \ifx0#2
    \node at (#1*\xmetaspacing+3.5*\xspacing,-0.4) {#3};
  \fi
  \ifx1#2
    \node at (#1*\xmetaspacing+3.5*\xspacing,1.4) {#3};
  \fi
}

\def\parsepos#1#2,#3{\def #1{#2*\xmetaspacing + #3*\xspacing}}
\newcommand\drawmatch[2]{
   \parsepos\posA#1
   \parsepos\posB#2
   \draw (\posA,0.1) -- (\posB,0.9);
}

\def\ind{0.11}
\def\drawpart#1,#2--#3:#4,#5--#6{
 \fill[gray] (#1*\xmetaspacing+#2*\xspacing-\ind,-\ind) -- (#1*\xmetaspacing+#2*\xspacing-\ind,+\ind) --
    (#4*\xmetaspacing+#5*\xspacing-\ind,1-\ind) -- (#4*\xmetaspacing+#5*\xspacing-\ind,1+\ind) --
    (#4*\xmetaspacing+#6*\xspacing+\ind,1+\ind) -- (#4*\xmetaspacing+#6*\xspacing+\ind,1-\ind) --
    (#1*\xmetaspacing+#3*\xspacing+\ind,\ind) -- (#1*\xmetaspacing+#3*\xspacing+\ind,-\ind) -- cycle;
}

\begin{document}
\renewcommand{\le}{\leqslant}
\renewcommand{\leq}{\leqslant}
\renewcommand{\ge}{\geqslant}
\renewcommand{\geq}{\geqslant}
\maketitle

\thispagestyle{empty}
\begin{abstract}

We consider codes over fixed alphabets against worst-case symbol
deletions.  For any fixed $k \ge 2$, we construct a
family of codes over alphabet of size $k$ with positive rate,
which allow efficient recovery from a worst-case deletion
fraction approaching $1-\frac{2}{k+\sqrt k}$. In particular, for binary
codes, we are able to recover a fraction of deletions approaching
$1/(\sqrt 2 +1)=\sqrt 2-1 \approx 0.414$. 
Previously, even non-constructively the largest deletion
fraction known to be correctable with positive rate was
$1-\Theta(1/\sqrt{k})$, and around $0.17$ for the binary case.

\smallskip 
Our result pins down the largest fraction of correctable deletions for
$k$-ary codes as $1-\Theta(1/k)$, since $1-1/k$ is an upper bound even
for the simpler model of erasures where the locations of the missing
symbols are known.

\smallskip
Closing the gap between $(\sqrt 2 -1)$ and $1/2$ for the limit of worst-case
deletions correctable by binary codes remains a tantalizing open
question.
\end{abstract}

\section{Introduction}

This work concerns error-correcting codes capable of correcting
{\em worst-case} deletions. Specifically, consider a fixed alphabet
$[k]\eqdef\{1,2,\dots,k\}$, and suppose we transmit a sequence of $n$
symbols from $[k]$ over a channel that can adversarially {\em delete}
an arbitrary fraction $p$ of symbols, resulting in a subsequence of
length $(1-p)n$ being received at the other end. The locations of the
deleted symbols are unknown to the receiver. The goal is to design a
code $C \subseteq [k]^n$ such that every $c \in C$ can be uniquely
recovered from any of its subsequences caused by up to $pn$
deletions. Equivalently, for $c \neq \tilde{c} \in C$, the length of the
longest common subsequence of $c,\tilde{c}$,  which we denote by $\LCS(c,\tilde{c})$, must be less than
$(1-p)n$.

In this work, we are interested in the question of correcting as large
a fraction $p$ of deletions as possible with codes of positive rate
(bounded away from $0$ for $n \to \infty$). That is, we would like
$|C| \ge \exp(\Omega_k(n))$ so that the code incurs only a constant
factor redundancy (this factor could depend on $k$, which we think of
as fixed).

Denote by $p^*(k)$ the limit superior of all $p \in [0,1]$ such that
there is a positive rate code family over alphabet $[k]$ that can
correct a fraction $p$ of deletions. The value of $p^*(k)$ is not known
for any value of $k$. Clearly, $p^*(k) \le 1-1/k$ --- indeed, one can
delete all but $n/k$ occurrences of the most frequent symbol in a
word to leave one of $k$ possible subsequences, and therefore only
trivial codes with $k$ codewords can correct a fraction $1-1/k$ of
deletions. This trivial limit remains the best known upper bound on
$p^*(k)$. We note that this upper bound holds even for the simpler
model of erasures where the locations of the missing symbols are known
at the receiver (this follows from the so-called Plotkin bound in
coding theory).

Whether the trivial upper bound $p^*(k) \le 1-1/k$ can be improved, or
whether there are in fact codes capable of correcting deletion
fractions approaching $1-1/k$ is an outstanding open question
concerning deletion codes and the combinatorics of longest common
subsequences. Perhaps the most notable of these is the $k=2$ (binary)
case.  The current best lower bound on $p^*(2)$ is around $0.17$. This
bound comes from the random code, in view of the fact that the
expected $\LCS$ of two random words in $\{0,1\}^n$ is at most
$0.8263n$~\cite{lueker-soda03}.  As the $\LCS$ of two random words in
$\{0,1\}^n$ is at least $0.788 n$, one cannot prove any lower bound on
$p^*(2)$ better than $0.22$ using the random code.  Kiwi, Loebl, and
Matou\v{s}ek~\cite{KLM} showed that, as $k\to\infty$, we have
$\mathbb{E}[\LCS(c,\tilde{c})]\sim \frac{2}{\sqrt{k}} n$ for two random words
$c,\tilde{c} \in[k]^n$.  This was used in \cite{GW-random15} to deduce $p^*(k)
\ge 1 - O(1/\sqrt{k})$.

The above discussion only dealt with the {\em existence} of deletion codes. Turning to explicit and efficiently decodable constructions,  Schulman and Zuckerman~\cite{SZ} constructed constant-rate binary codes which are efficiently decodable from a {\em small} constant fraction of worst-case deletions. This was improved in \cite{GW-random15}; in the new codes, the rate approaches~$1$.
Specifically, it was shown that one can correct a fraction $\zeta > 0$ of deletions with rate about $1-O(\sqrt{\zeta})$.
In terms of correcting a larger fraction of deletions, codes that are efficiently decodable from a fraction $1-\gamma$ of errors over a $\mathrm{poly}(1/\gamma)$ sized alphabet were also given in \cite{GW-random15}.

Our focus in this work is exclusively on the worst-case model of deletions. For random deletions, it is known that reliable communication at positive rate is possible for deletion fractions approaching $1$ even in the binary case. We refer the reader interested in coding against random deletions to the survey by Mitzenmacher~\cite{M-survey}.

\subsection{Our results}
Here we state our results informally, omitting the precise computational efficiency guarantees, and omitting
the important technical properties of constructed codes related to the ``span'' of common
subsequences (see Section~\ref{sec:prelim} for the definition). The precise statements are in Subsection~\ref{subsec:efficient}
and in Section~\ref{sec:decoding}.

Our first result is a construction of codes which are combinatorially capable of correcting a larger fraction of deletions than was previously known to be possible.

\begin{theorem}[Informal]
For all integers $k \ge 2$, $p^*(k) \ge 1-\frac{2}{k+\sqrt k}$.
Furthermore, for any desired $\eps > 0$, there is an efficiently constructible family of $k$-ary codes of rate $r(k,\eps)>0$ such that the $\LCS$ of any two distinct codewords is less than fraction $\frac{2}{k+\sqrt k}+\eps$ of the code length.
In particular, there are explicit binary codes that can correct a fraction $(\sqrt 2 - 1 -\eps) > 0.414-\eps$ of deletions, for any fixed $\eps > 0$. 
\end{theorem}

Note that, together with the trivial upper bound $p^*(k) \le 1-1/k$, the result pins down the asymptotics of $1-p^*(k)$ to $\Theta(1/k)$ as $k\to\infty$. Interestingly, our result shows that deletions are easier to correct than errors (for worst-case models), as one cannot correct a fraction $1/4$ of worst-case errors with positive rate.

In our second result we construct codes with the above guarantee together with an efficient algorithm to recover from deletions:

\begin{theorem}[Informal]
For any integer $k \ge 2$ and any $\eps > 0$, there is an efficiently constructible  family of $k$-ary codes of rate $r(k,\eps) > 0$ that can be decoded in polynomial (in fact near-linear)  time from a fraction $1 -\frac{2}{k+\sqrt k} -\eps$ of deletions.
\end{theorem}

\subsection{Our techniques}

All our results are based on code concatenations, which use an outer
code over a large alphabet with desirable properties, and then further
encode the codeword symbols by a judicious inner code.  The inner code
comes in two variants, one clean and simpler form and then a dirty
more complicated form giving a slightly more involved and better
bounds.  For simplicity let us here describe the clean construction
which when analyzed gives the slightly worse bound $1-\frac {2}{k+1}$
as compared to $1-\frac {2}{k+\sqrt k}$. This weaker bound appears in
the preliminary conference version~\cite{BG-soda16} of this paper.

The innermost code consists of words of the form
$(\1^A\2^A\ldots\textsf{k}^A)^{L/A}$ for integers $A,L$ with $A$ dividing $L$, where $\alpha^A$ stands for the letter $\alpha$ repeated $A$ times.  Informally, we think of these
words as oscillating with amplitude $A$ (this can be made precise via
Fourier transform for example, but we won't need it in our
analysis). The crucial property, that was observed in
\cite{bukh_zhou}, is that two such words have a long common
subsequence only if their amplitudes are close. This property was
also exploited in \cite{bukh_ma} to show a certain weak limitation of
deletion codes, namely that in any set of $t \ge k+2$ words in
$[k]^n$, some two of them have an LCS at least $\frac{n}{k} + c(k,t)
n^{1-1/(t-k-2)}$.

The effective use of these codes as inner codes in a concatenation
scheme relies on a property stronger than absence of long common
subsequences between codewords. Informally, the property amounts to
absence of long common subsequences between subwords of codewords. For
the precise notion, consult the definition of a {\em span} in the next
section and the statement of Theorem~\ref{thm:alphabet-reduction-clean} in
the following section. Using this, we are able to show that if the
outer code has a small $\LCS$ value, then the LCS of the concatenated
code approaches a fraction $\frac{2}{k+1}$ of the block length.

For the outer code, the simplest choice is the random code. This gives
the existential result (Theorem~\ref{thm:main-existential}).  Using the
explicit construction of codes to correct a large fraction of
deletions over fixed alphabets from \cite{GW-random15} gives us a
polynomial (in fact near-linear) time deterministic construction (Theorem~\ref{thm:deletion-codes-constructive}).
While the outer code from \cite{GW-random15} is also efficiently
decodable from deletions, it is not clear how to exploit this to
decode the concatenated code efficiently.

To obtain codes that are also efficiently decodable, we employ another
level of concatenation, using Reed--Solomon codes at the outermost
level, and the above explicit concatenated code itself as the inner
code. The combinatorial LCS property of these codes is established
similarly, and is in fact easier, as we may assume (by indexing each
position) that all symbols in an outer codeword are distinct, and
therefore the corresponding inner codewords are distinct.  To decode
the resulting concatenated code, we try to decode the inner code (by
brute-force) for many different contiguous subwords of the received
subsequence. A small fraction of these are guaranteed to succeed in
producing the correct Reed--Solomon symbol. The decoding is then
completed via \emph{list decoding} of Reed--Solomon codes. The
approach here is inspired by the algorithm for list decoding binary
codes from a deletion fraction approaching $1/2$ in
\cite{GW-random15}. Our goal here is to recover the correct message
     {\em uniquely}, but by virtue of the combinatorial guarantee,
     there can be at most one codeword with the received word as a
     subsequence, so we can go over the (short) list and identify the
     correct codeword. Note that list decoding is used as an
     intermediate algorithmic primitive even though our goal is unique
     decoding; this is similar to \cite{GI-soda04} that
     gave an algorithm to decode certain low-rate concatenated codes
     up to half the Gilbert--Varshamov bound via a list decoding
     approach.

\section{Preliminaries}\label{sec:prelim}
A \emph{word} is a sequence of symbols from a finite alphabet.
For the problems of this paper, only the size of the alphabet and the length of the word are important.
So, we will often use $[k]$ for a canonical $k$-letter alphabet,
and consider the words indexed by $[n]$. In this case, the set of words of length $n$ over alphabet $[k]$
will be denoted $[k]^n$. We treat symbols in a word as \emph{distinguishable}. So, if $x$ denotes
the second $\1$ in the word $\2\1\0\1\1$ and we delete the subword $\1\0$, the variable $x$
now refers to the first $\1$ in the word $\2\1\1$.

Below we define some terminology about subsequences that we will use throughout the paper:
\begin{itemize}

\item 
A \emph{subsequence} in a word $w$ is any word obtained from $w$ by deleting
one or more symbols.  In contrast, a \emph{subword} is a subsequence
made of several consecutive symbols of~$w$.  

\item The \emph{span} of a
subsequence $w'$ in a word $w$ is the length of the smallest subword
containing the subsequence. We denote it by $\wspan_w w'$, or simply
by $\wspan w'$ when no ambiguity can arise.

\item
A \emph{common subsequence} between words $w_1$ and $w_2$ is a pair $(w_1',w_2')$ of
subsequences $w_1'$ in $w_1$ and $w_2'$ in~$w_2$ that are equal as
words, i.e., $\len w_1'=\len w_2'$ and the $i$'th symbols of $w_1'$ and $w_2'$ are equal for each $i$, $1 \le i \le \len w_1'$. 

\item 
For words $w_1,w_2$, we denote by $\LCS(w_1,w_2)$ the length of the
\emph{longest common subsequence} of $w_1$ and $w_2$, i.e., the
largest $j$ for which there is a common subsequence between $w_1$ and
$w_2$ of length $j$.
\end{itemize}

A \emph{code} $C$ of block length $n$ over the alphabet $[k]$ is simply a subset of $[k]^n$. We will also call such codes as $k$-ary codes, with binary codes referring to the $k=2$ case. The \emph{rate} of $C$ equals $\frac{\log |C|}{n \log k}$.

For a code $C \subseteq [k]^n$, its {\em LCS value} is defined as 
\[ \LCS(C)\eqdef\max_{c_1 \neq c_2 \in C} \LCS(c_1,c_2) \ .\]
Note that a code $C \subseteq [k]^n$ is capable of recovering from $t$
worst-case deletions if and only if $\LCS(C) < n-t$.

We define the \emph{span of a common subsequence} $(w_1',w_2')$ of words $w_1$ and $w_2$ as
\[
  \wspan (w_1',w_2')\eqdef\wspan_{w_1} w_1'+\wspan_{w_2} w_2'.
\]
The span will play an important role in our analysis of $\LCS(C)$ of
the codes $C$ we construct, by virtue of the fact that if $\wspan
(w_1',w_2') \ge b \cdot \len w_1'$ for every common subsequence of
$w_1,w_2 \in [k]^n$, then $\LCS(w_1,w_2) \le \frac{2n}{b}$.  Our
result will be based on a construction for which we can take $b
\approx k+\sqrt k$ for long enough common subsequences of any distinct pair
of codewords.

\smallskip \noindent {\bf Concatenated codes.} Our results heavily use the simple but useful idea of code concatenation. Given an {\em outer} code $C_{\rm out} \subseteq [Q]^n$, and an injective map
$\tau \colon [Q] \to [q]^m$ defining the encoding function of an {\em inner} code $C_{\rm in}$, the concatenated code $C_{\rm concat} \subseteq [q]^{nm}$ is obtained by composing these codes as follows. If $(c_1,c_2,\dots,c_n) \in [Q]^n$ is a codeword of $C_{\rm out}$, the corresponding codeword in $C_{\rm concat}$ is $(\tau(c_1),\dots,\tau(c_n)) \in [q]^{nm}$. The words $\tau(c_i) \in C_{\rm in}$ will be referred to as the {\em inner blocks} of the concatenated codeword, with the $i$'th block corresponding to the $i$'th outer codeword symbol.

\section{Alphabet reduction for deletion codes}

Fix $k$ to be the alphabet size of the desired deletion code. We shall
show how to turn words over $K$-letter alphabet, for $K \gg k$,
without large common subsequence into words over $k$-letter alphabet
without large common subsequence.  More specifically, for any $\eps >
0$ and large enough integer $K = K(\eps)$, we give a method to
transform a deletion code $C_1\subseteq [K]^n$ with $\LCS(C_1) \ll
\eps n$ into a deletion code $C_2 \subseteq [k]^N$ with $\LCS(C_2) \le
\bigl(\frac{2}{k+\sqrt k}+\eps\bigr) N$. The transformation lets us
transform a {\em crude} dependence between the alphabet size of the
code $C_1$ and its LCS value (i.e., between $K$ and $\eps$), into a
{\em quantitatively strong} one, namely $\LCS(C_2) \approx
\frac{2}{k+\sqrt k} N$.  The code $C_2$ will in fact be obtained by concatenating $C_1$ with an inner $k$-ary code with $K$ codewords, and therefore has the same cardinality as $C_1$.
The block length $N$ of $C_2$ will be much larger than $n$, but the ratio
$N/n$ will be bounded as a function of $k,K$, and $\eps$. The rate of
$C_2$ will thus only be a constant factor smaller than that of $C_1$.

Specifically, we prove the following.
\begin{theorem}
\label{thm:alphabet-reduction-dirty}
Let $C_1 \subseteq [K]^n$ be a code with $\LCS(C_1) = \gamma n$, and
let $k \ge 2$ be an integer. Then there exists an integer $T =
T(K,\gamma,k)$ satisfying $T \le O((2k/\gamma)^{2K+2})$, and an injective map
$\tau\colon [K] \to [k]^T$ such that the code $C_2 \subseteq [k]^{N}$ for
$N=nT$ obtained by replacing each symbol in codewords of $C_1$ by its
image under $\tau$ has the following property: if $s$ is a common
subsequence between two distinct codewords $c,\tilde{c} \in C_2$, then
\begin{equation}
\label{eq:dirty-alphabet-reduction}
\wspan s \ge (k+\sqrt k) \len s - 5 \gamma k N \ . 
\end{equation}
In particular, since $\wspan s \le 2N$, we have $\LCS(C_2) \le \left( \frac{2+5 \gamma k}{k+\sqrt k} \right) N < \left( \frac{2}{k+\sqrt k} + 5 \gamma \right) N$.
\end{theorem}

Thus, one can construct codes over a size $k$ alphabet with LCS value
approaching $\frac{2}{k+\sqrt k}$ by starting with an outer code with LCS value
$\gamma \to 0$ over any fixed size alphabet, and concatenating it with
a constant-sized map. The span property will be useful
in concatenated schemes to get longer, efficiently decodable
codes.

The key to the above construction is the inner map, which 
come in two variants, one ``clean'' and one ``dirty'' form.
The former is simpler to describe and we choose to do this first.

\subsection{The clean construction}

The aim of the clean construction is to prove the following:
\begin{theorem}
\label{thm:alphabet-reduction-clean}
Let $C_1 \subseteq [K]^n$ be a code with $\LCS(C_1) = \gamma n$, and
let $k \ge 2$ be an integer. Then there exists an integer $T =
T(K,\gamma,k)$ satisfying $T \le 32 \cdot (2k/\gamma)^K$, and an injective map
$\tau\colon [K] \to [k]^T$ such that the code $C_2 \subseteq [k]^{N}$ for
$N=nT$ obtained by replacing each symbol in codewords of $C_1$ by its
image under $\tau$ has the following property: if $s$ is a common
subsequence between two distinct codewords $c,\tilde{c} \in C_2$, then
\[ \wspan s \ge (k+1) \len s - 4 \gamma k N \ . \]
In particular, since $\wspan s \le 2N$, we have $\LCS(C_2) \le \left( \frac{2+4\gamma k}{k+1} \right) N < \left( \frac{2}{k+1} + 4 \gamma \right) N$.
\end{theorem}

We start by describing the way to encode symbols from the alphabet $[K]$ as
words over $[k]$ that underlies
Theorem~\ref{thm:alphabet-reduction-clean}.  Let $L$ be constant to be
chosen later. For an integer $A$ dividing $L$, define the word of ``amplitude $A$'' to be
\begin{equation}
\label{eq:freqs}
  f_A\eqdef (\1^A\2^A\ldots\textsf{k}^A)^{L/A}.
\end{equation}
where $\alpha^A$ stands for the letter $\alpha$ repeated $A$ times.
The crucial property of these words is that $f_A$ and $f_B$ have no
long common subsequence if $B/A$ is large (or small); for the proof see one of \cite{bukh_zhou,bukh_ma}.
In the present work, we will need
a more general ``asymmetric'' version of this observation --- we will need to analyze
common subsequences in subwords of $f_A$ and $f_B$ (which may be of different lengths)

Let $R\geq k$ be an integer to be chosen later. For
a word $w$ over alphabet $[K]$ denote by $\hat{w}$ the word obtained
from $w$ via the substitution
\begin{equation}
\label{eq:K-to-k}
  l\in[K]\quad\mapsto\quad f_{R^{l-1}}.
\end{equation}
to each symbol of $w$.  Note that $\len \hat{w}=kL\len w$. If a symbol
$x\in \hat{w}$ is obtained by expanding symbol $y\in w$, then we say
that $y$ is a \emph{parent} of $x$.

\subsubsection{Analysis of clean construction}
\begin{lemma}\label{lemma:basicspan}
For a natural number $P$, let $f^{\infty}_A$ be the (infinite) word
\[
  (\1^A\2^A\ldots\textsf{k}^A)^*
\]
Let $A,B$, where $kA \leq B$ be natural numbers, and 
suppose $s=(w_1',w_2')$ is a common
subsequence between $f^{\infty}_A$ and $f^{\infty}_B$. Then
\[
  \wspan s\geq \left(k+1-\frac{kA}{B}\right)\len s-2(A+B).
\]
\end{lemma}
\begin{proof}
The words $f^{\infty}_A$ and $f^{\infty}_B$ are concatenations of
\emph{chunks}, which are subwords of the form $l^A$ and $l^B$
respectively. A chunk in $f^{\infty}_A$ is \emph{spanned} by
subsequence $w_1'$ if the span of $w_1'$ contains at least one symbol
of the chunk. Similarly, we define chunks spanned by $w_2'$ in
$f^{\infty}_B$. We will estimate how many chunks are spanned by $w_1'$
and by $w_2'$.

As a word, a common subsequence is of the form
$k_1^{p_1}k_2^{p_2}\dotsb k_s^{p_s}$ where $k_l\neq k_{l+1}$ and the
exponents are positive.  The subsequence $k_l^{p_l}$ spans at least
$k\left\lceil \frac{p_l-A}{A}\right\rceil+1$ chunks in
$f^{\infty}_A$. Similarly, $k_l^{p_l}$ spans at least $k\left\lceil
\frac{p_l-B}{B}\right\rceil+1$ chunks in $f^{\infty}_B$. Therefore the
total number of symbols in chunks spanned by $k_l^{p_l}$ in both
$f^{\infty}_A$ and in $f^{\infty}_B$ is at least
\[
  \phi(p_l)\eqdef A\left(k\left\lceil
  \frac{p_l-A}{A}\right\rceil+1\right)+B\left(k\left\lceil
  \frac{p_l-B}{B}\right\rceil+1\right)
\]
We then estimate $\phi(p_l)$ according to whether $p_l\leq B$:
\[
  \phi(p_l)\geq
\begin{cases}
k(p_l-A)+B&\text{if }p_l\leq B,\\[0.6em]
k(p_l-A)+k(p_l-B)+B&\text{if }p_l>B.
\end{cases}
\]
In both cases we have
\[
  \phi(p_l)\geq \left(k+1-\frac{kA}{B}\right)p_l.
\]

Note that the chunks spanned by $k_l^{p_l}$ are distinct from chunks
spanned by $k_{l'}^{p_{l'}}$ for $l\neq l'$.  So, the total number of
symbols in all chunks spanned by subsequence $s$ in both
$f^{\infty}_A$ and $f^{\infty}_B$ is least
\[
  \sum_l \phi(p_l)\geq \left(k+1-\frac{kA}{B}\right)\len s.
\]
The total span of $s$ might be smaller since the first and the last
chunks in each of $f^{\infty}_A$ and $f^{\infty}_B$ might not be fully
spanned. Subtracting $2(A+B)$ to account for that gives the stated
result.
\end{proof}

Let $(w_1',w_2')$ be a common subsequence between $\hat{w}_1$ and
$\hat{w}_2$. We say that the $i$'th symbol in $(w_1',w_2')$ is
\emph{well-matched} if the parents of $w_1'[i]$ and of $w_2'[i]$ are the same letter of $[K]$.
A common subsequence is \emph{badly-matched} if none of its
symbols are well-matched; see Figure 1 below for an example.

\def\subseqpic{

\drawmetasymbol{0}{1}{1}

\drawmetasymbol{1}{1}{3}

\drawmetasymbol{2}{1}{3}

\drawmetasymbol{3}{1}{2}

\drawmetasymbol{4}{1}{1}

\drawmetasymbol{0}{0}{2}

\drawmetasymbol{1}{0}{2}

\drawmetasymbol{2}{0}{1}

\drawmetasymbol{3}{0}{3}

\drawmetasymbol{4}{0}{1}

\drawmatch{0,0}{0,1}

\drawmatch{0,1}{0,3}

\drawmatch{0,2}{0,4}

\drawmatch{0,3}{0,5}

\drawmatch{0,4}{1,0}

\drawmatch{0,6}{1,1}

\drawmatch{1,0}{1,2}

\drawmatch{1,1}{1,4}

\drawmatch{1,2}{1,5}

\drawmatch{1,4}{1,6}

\drawmatch{2,4}{1,7}

\drawmatch{2,5}{2,1}

\drawmatch{2,6}{2,3}

\drawmatch{2,7}{2,5}

\drawmatch{3,0}{3,0}

\drawmatch{3,2}{3,1}

\drawmatch{3,3}{3,2}

\drawmatch{3,4}{3,4}

\drawmatch{3,5}{3,5}

\drawmatch{3,6}{4,0}

\drawmatch{3,7}{4,4}}

\begin{tikzpicture}

\subseqpic

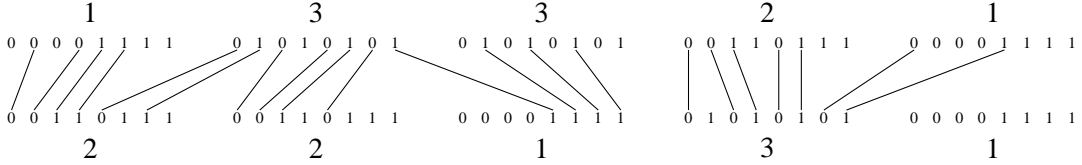
\node at (2*\xmetaspacing+3.5*\xspacing,-1.0) {Figure 1: A badly-matched common subsequence between $w_1'$ and $w_2'$ for $w_1=\1\3\3\2\1$ and $w_2=\2\2\1\3\1$};

\end{tikzpicture}

\begin{lemma}\label{lem:badly}
Suppose $w_1,w_2$ are words over alphabet $[K]$ and $s=(w_1',w_2')$ is a badly-matched common subsequence between
$\hat{w}_1$ and $\hat{w}_2$ as defined in \eqref{eq:K-to-k}. Then
\[
  \wspan w_1'+\wspan w_2'\geq \left(k+1-\frac{k}{R}-\frac{8R^{K-1}}{L}\right)\len s-16R^{K-1}.
\]
\end{lemma}
\begin{proof}
We subdivide the common subsequence $s$ into subsequences
$s_1,\dotsc,s_r$ such that, for each $i=1,\dotsc,r$ and each $j=1,2$,
the symbols matched by $s_i$ in $w_j'$ belong to the expansion of the
same symbol in $w_j$. We choose the subdivision to be a coarsest one
with this property (see Figure 2 below for an example).  That implies that pairs of symbols of $w_1$ and
$w_2$ matched by $s_i$ and by $s_{i+1}$ are different. In particular,
expansions of at least $r-4$ symbols of $w_1$ and $w_2$ (except possibly the expansions of the leftmost and rightmost symbols of each of them) are fully
contained in the spans of $w_1'$ and $w_2'$. Therefore, we have
\[
  Lk(r-4)\leq \wspan s.
\]
Since $(w_1',w_2')$ is badly-matched, by the preceding lemma we then have
\begin{align*}
  \wspan s&\geq \left(k+1-\frac{k}{R}\right)\len s-4r R^{K-1} \geq \left(k+1-\frac{k}{R}\right)\len s-4R^{K-1}\left(\frac{\wspan s}{Lk}+4\right).
\end{align*}
The lemma then follows from the collecting together the two terms involving $\wspan w_1'+\wspan w_2'$, and
then dividing by $1+4R^{K-1}/Lk$.
\end{proof}

\begin{tikzpicture}

\drawpart0,0--3:0,1--5
\drawpart0,4--6:1,0--1
\drawpart1,0--4:1,2--6
\drawpart2,4--4:1,7--7
\drawpart2,5--7:2,1--5
\drawpart3,0--5:3,0--5
\drawpart3,6--7:4,0--4

\subseqpic

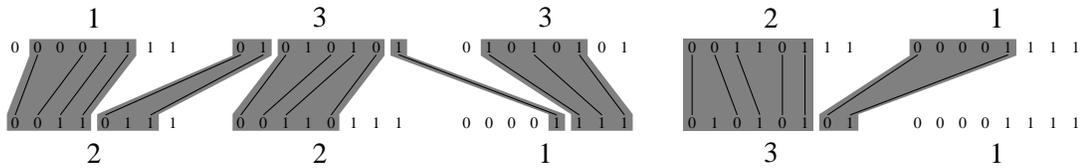
\node at (2*\xmetaspacing+3.5*\xspacing,-1.0) {Figure 2: Partition of the common subsequence from Figure 1 into subsequence as in the proof of Lemma~\ref{lem:badly}};

\end{tikzpicture}

The next step is to drop the assumption in Lemma~\ref{lem:badly} that the common subsequence is badly-matched.
By doing so we incur an error term involving $\LCS(w_1,w_2)$.
\begin{lemma}
\label{lem:notbadly}
Suppose $w_1,w_2$ are words over alphabet $[K]$ and $s=(w_1',w_2')$ is a common subsequence between
$\hat{w}_1$ and $\hat{w}_2$. Then
\[
  \wspan s\geq \left(k+1-\frac{k}{R}-\frac{8R^{K-1}}{L}\right)\len s-2Lk(k+1)\cdot \LCS(w_1,w_2)-16R^{K-1}.
\]
\end{lemma}
\begin{proof}
Without loss, the subsequence $s$ is locally optimal, i.e., every
alteration of $s$ that increases $\len s$ also increases $\wspan
s$. Define an auxiliary bipartite graph $G$ whose two parts are the
symbols in $w_1$ and the symbols in $w_2$. For each well-matched
symbol in $s$ we join the parent symbols in $w_1$ and $w_2$ by an
edge.

We may assume that each vertex in $G$ has degree at most~$2$. Indeed,
suppose a symbol $x\in w_1$ is adjacent to three symbols $y_1,y_2,y_3\in
w_2$ with $y_2$ being in between $y_1$ and $y_3$. Then we alter $s$ by
first removing all matches between $x$ and $y_1,y_2,y_3$, and then
completely matching $x$ with $y_2$. The alteration does not increase
$\wspan s$, and the result is a common subsequence that is at least as
long as $s$, and whose auxiliary graph has fewer edges. We can then
repeat this process until no vertex has degree exceeding $2$.

Consider a maximum-sized matching in $G$. On one hand, it has at most
$\LCS(w_1,w_2)$ edges. On the other hand, since the maximum degree of $G$
is at most $2$, the maximum-sized matching has at least $\abs{E(G)}/2$
edges. Hence, $\abs{E(G)}\leq 2\LCS(w_1,w_2)$.

Remove from $s$ all well-matched symbols to obtain a common
subsequence $s'$.  The new subsequence satisfies
\[
  \len s'\geq \len s-Lk\cdot \abs{E(G)}\geq \len s-2Lk\cdot \LCS(w_1,w_2).
\]
It is also clear that $s'$ is a badly-matched common subsequence. From
the previous lemma
\[
  \wspan s'\geq \left(k+1-\frac{k}{R}-\frac{8R^{K-1}}{L}\right)\len s-2Lk(k+1)\cdot \LCS(w_1,w_2)-16R^{K-1}.
\]
Since $\wspan s\geq \wspan s'$, the lemma follows.
\end{proof}

We are now ready to prove Theorem~\ref{thm:alphabet-reduction-clean} by picking parameters suitably.

\begin{proof}[Proof of Theorem~\ref{thm:alphabet-reduction-clean}]
Recall that we are starting with a code $C_1 \subseteq [K]^n$ with $\LCS(C_1) = \gamma n$. Given $\eps > 0$ and an integer $k \ge 2$, pick parameters
\[ R = \left\lceil \frac{2k}{\gamma}\right\rceil \quad \mbox{and} \quad L = 16 R^{K-1}\left\lceil \frac{1}{\gamma }\right\rceil \]
in the construction \eqref{eq:freqs} and \eqref{eq:K-to-k}. Define $T = kL$ and $\tau \colon [K] \to [k]^T$ as $\tau(l) = f_{R^{l-1}}$, and let $C_2 \subseteq [k]^N$, where $N=nkL$, be the code obtained as in the statement of Theorem~\ref{thm:alphabet-reduction-clean}. Note that $T \le 32 \cdot (2k/\gamma)^K$ by our choice of parameters.

By Lemma~\ref{lem:notbadly}, we can conclude that any common subsequence $s$ of two distinct codewords of $C_2$ satisfies
\[ \wspan s \ge (k+1 - \gamma) \len s - 2 (k+1) \gamma N - \gamma N \ . \]
Since $\len s \le N$ and $k \ge 2$, the right hand side is at least $(k+1) \len s - 4 k \gamma N$, as desired.
\end{proof}

\begin{remark}[Bottleneck for analysis]
We now explain why the analysis in Theorem~\ref{thm:alphabet-reduction-clean} is limited to proving correctability of a $1/3$ fraction of deletions for binary codes (a similar argument holds for larger alphabet size $k$). Imagine subwords of length $3$ of $w_1,w_2 \in [K]^n$ of the form $abc$ and $def$ respectively, where $d > a,b$ and $c > e,f$. Then the word $f_{R^{d-1}}$  can be matched fully with $f_{R^{a-1}} f_{R^{b-1}}$ (because the latter strings oscillate at a higher frequency that $f_{R^{d-1}}$), and similarly $f_{R^{c-1}}$ can be matched fully with $f_{R^{e-1}} f_{R^{f-1}}$. Thus we can find a common subsequence of length $4L$ between the encoded bit strings $f_{R^{a-1}} f_{R^{b-1}} f_{R^{c-1}} \in [2]^{6L}$ and $f_{R^{d-1}} f_{R^{e-1}} f_{R^{f-1}}\in [2]^{6L}$, even if $abc$ and $def$ share no common subsequence. 
\end{remark}

\subsection{Dirty construction}

We now turn to the more complicated ``dirty'' construction in which
small runs of dirt are interspersed in the long runs of a single
symbol from the clean construction.

\subsubsection{Dirty construction, binary case}

To convey the intuition for the dirty construction let us look more
closely at what happened in the binary case.  We were looking
for subsequences of 
\[
f^{\infty}_A =  (\1^A\2^A)^*
\]
and
\[
f^{\infty}_B =  (\1^B\2^B)^*
\]
where both $A$ and $B$ are large numbers but $B$ is much larger than $A$.
We are interested in subsequences with small span.  Looking more closely
at the proof of Lemma~\ref{lemma:basicspan} we see that such
subsequences are
obtained by taking every symbol of $f^{\infty}_B$ and discarding 
essentially half the symbols of $f^{\infty}_A$ as to not interrupt the
very long runs in  $f^{\infty}_B$.  Now suppose we introduce some ``dirt'' in
$f^{\infty}_B$ by introducing, in the very long stretches of 1's, some infrequent
2's, say a 2 every 10th symbol (and similarly some infrequent 1's in
the long stretches of 2's).  Then, during construction of the LCS, 
when running into such 
a sporadic 2 we can either try to include it or discard it.
As $A$ is a large number it is easy to see that while we
are matching a 1-segment of $f^{\infty}_A$ we cannot profit by matching
the sporadic 2's.  It is also not difficult to see that while
passing through a 2-segment of $f^{\infty}_A$ it is not profitable
to match more than one sporadic 2 as matching two consecutive sporadic 2's
forces us to drop the ten 1's in between the two matched 2's
in $f^{\infty}_B$.
The net effect is that introducing some dirt hardly enables us to
expand the LCS but does increase the span.  We need to introduce dirt
in all codewords and it should not look too similar in any two codewords.
The way to achieve this is by introducing such dirty runs of different but short
lengths in all codewords. Let us turn to a more formal description.  

For the sake of readability
we below assume that some real numbers defined are integers.
Rounding these numbers to the closest integer only introduces
lower order term errors.  It is also not difficult to
see that we can pick parameters such that all numbers are indeed integers.

Let $c$ be a such that $0 \leq c < \sqrt 2- 1$.  The reason for the upper limit on $c$ will be clarified in Remark~\ref{rem:dirty} after the analysis. 
We define ``$M$ dirty ones at
amplitude $a$'' be the string
\[
(1^a 2^{ca})^{M/(1+c)a}
\]
and let us write this as $1_{M,a}$ leaving $c$ implicit.  We
have an analogous string $2_{M,a}$ and we allow $M=\infty$ with the
natural interpretation. 
Remember that in our clean solution, $i$ was coded by 
\[
f_{R^{i-1}}=(1^{R^{i-1}} 2^{R^{i-1}}) ^{L/R^{i-1}}.
\] 
In the dirty construction we replace this by
\begin{equation}
\label{eq:gi}
g_i=(1_{R^{K+1+i},R^{K-i} },2_{R^{K+1+i},R^{K-i} })^{L/R^{K+1+i}},
\end{equation}
where $R$ is an integer that can be written on the form $(1+c)t$ for an
integer $t$, and 
\begin{equation}
L = R^{2K+1} \ .
\end{equation}

We introduce dirt where the amplitude of the dirt decreases
with $i$.  We call a string of the form $j_{R^{K+1+i},R^{K-i}}$ as a
{\bf {\em segment}} of $g_i$.
The reason for the general length increase by a 
factor $R^{k+1}$ is to accommodate for dirt of frequencies that
are well separated.
\begin{lemma}\label{lemma:inftyw1}
Let $w_1$ be the string
$1_{\infty,a}$ (or $2_{\infty,a}$) and let $s$ be a subsequence of 
$w_2=(1_{b_1,b_2},2_{b_1,b_2})$, then
\[
  \wspan_{w_1}{s} +2b_1 \geq (3+c)\len s-\frac {4ab_1}{b_2}
\]
\end{lemma}
\begin{proof}
As $w_2$ is symmetric in 1 and 2 we can assume that $w_1=1_{\infty,a}$.
Note that $w_2$ consists of substrings of the
form $1^{b_2}, 1^{cb_2}, 2^{b_2},$ and $2^{cb_2}$ and $b_1/(1+c)b_2$ copies
of each.   For the $i$'th subword of ones (ignoring if it 
is of length $b_2$ or $cb_2$), let us assume that $s_i$ 1's are
contained in $s$.  The the span of this subsequence in $w_1$ is
at least $(s_i/a-1) (1+c)a$.  Similarly if the
$i$'th string of 2's contain $t_i$ symbols from $s$ then
its span in $w_1$ is at least $(t_i/(ca)-1) (1+c)a$.
Summing these inequalities, if $S$ is the total number of 1's in 
$s$ and $\tilde{S}$ is the number of 2's, then the span of $s$ in $w_1$ is at least
$$
(1+c) S + (1+c)\tilde{S}/c- \frac {4ab_1}{b_2} \ ,
$$
where the last term comes because we lose $(1+c)a$ in the span for each substring of identical symbols in $w_2$ and there are $4b_1/(1+c)b_2$ such substrings.

As the length of $s$ is $S+\tilde{S}$ it is sufficient to establish that
\begin{equation}\label{stineq}
(1+c) S + (1+c)\tilde{S}/c + 2b_1 \geq (3+c) (S+\tilde{S}).
\end{equation}
We know that both $S$ and $\tilde{S}$ are in the range $[0, b_1]$. 
Since $0 \leq c < \sqrt 2-1$ we have $(1+c)/c > (3+c)$
and thus it is sufficient to establish (\ref{stineq}) for $\tilde{S}=0$, but
in this case it follows from $S \leq b_1$.
\end{proof}

The above lemma is the main ingredient in establishing the
the following lemma.

\begin{lemma}\label{lemma:gigjdirty}
Let $s$ be a subsequence of $g_i$ and $g_j$ for $i< j$, then, provided
$R\geq 10$, 
\[
  (1+\frac 2R)\wspan_{g_i}{s} +  \wspan_{g_j}{s} \geq (3+c)\len s-\frac {10L}{R}
\]
\end{lemma}
\begin{proof}
We have that $g_i$ consists of $L/R^{K+1+i}$
substrings of each of the form
$$1_{R^{K+1+i},R^{K-i}},2_{R^{K+1+i},R^{K-i}}.$$
Now partition $s$ into substrings $s^{(k)}$ according to how it intersects
these substrings of $g_i$.  The number of such strings
is at most $2+ (\wspan_{g_i}{s})/(2R^{K+1+i})$.
We want to apply Lemma~\ref{lemma:inftyw1} 
and we need to address the fact that each $s^{(k)}$ might intersect more than
one segment of $g_j$ (recall that a segment of $g_j$ is a substring of the form $1_{R^{K+1+j},R^{K-j}}$ or $2_{R^{K+1+j},R^{K-j}}$).  As $g_j$ only has $2L/R^{K+1+j}$
different segments, by refining the partition 
slightly we can obtain substrings
$s^{(k)}$ for $k=1, \ldots, p$ 
with $p \leq 2+(\wspan_{g_i}{s})/(2R^{K+1+i})+2L/R^{K+1+j}$,
where each $s^{(k)}$ satisfies the hypothesis of Lemma~\ref{lemma:inftyw1} with
$a=R^{K-j}$, $b_1=R^{K+i+1}$ and $b_2=R^{K-i}$. We therefore obtain 
the inequality 
\begin{equation}\label{eq:sk}
  \wspan_{g_j}{s^{(k)}} +2R^{K+i+1} \geq (3+c)\len s^{(k)}- 4R^{i-j}R^{K+i+1}.
\end{equation}
We have a total of $p$ inequalities and as  $\wspan_{g_j}{s} \geq
\sum_k \wspan_{g_j}{s^{(k)}} $ and $\len s =\sum_k \len s^{(k)}$,
summing (\ref{eq:sk}) for the $p$ values of $k$ gives
\[
  \wspan_{g_j}{s} +2 p R^{K+i+1} \geq (3+c)\len s- 4p R^{i-j}R^{K+i+1}.
\]
Now as 
$p \leq 2+\wspan_{g_i}{s}/(2R^{K+1+i})+2L/R^{K+1+j}$
we can conclude that
\[
\wspan_{g_j}{s} +  (1+2R^{i-j}) \wspan_{g_i}{s} \geq (3+c)\len s- (4R^{K+i+1}+4LR^{i-j}+8R^{i-j} R^{K+i+1}+8R^{2(i-j)}L)
\]
and using $R\geq 10$, $R^{K+i+1}\leq \frac LR$, and $i < j$, the lemma follows.
\end{proof}

Let us slightly abuse notation and in this section let
$\hat{w}$ the word obtained  from a word $w$ via the substitution
\begin{equation}
\label{eq:K-to-k-dirty}
  l\in[K]\quad\mapsto\quad g_{l}
\end{equation}
to each symbol of $w$ as opposed to (\ref{eq:K-to-k}).  As Lemma~\ref{lemma:gigjdirty} tells us that subsequences of codings of unequal symbols
have a large span, we have the following analog of Lemma~\ref{lem:badly}.

\begin{lemma}\label{lem:badly-dirty}
Suppose $w_1,w_2$ are words over alphabet $[K]$ and $s=(w_1',w_2')$ is a badly-matched common subsequence between
$\hat{w}_1$ and $\hat{w}_2$ as defined in \eqref{eq:K-to-k-dirty}. Then
\[
  \wspan w_1'+\wspan w_2'\geq \left(3+c-\frac {28}{R}\right)\len s-\frac {40L}{R}.
\]
\end{lemma}
\begin{proof}
We use the same subdivision as in the proof Lemma~\ref{lem:badly}.
We have
\[
  2L(r-4)\leq \wspan s.
\]
Since $(w_1',w_2')$ is badly-matched, by the preceding lemma we then have
\begin{align*}
\left(1+\frac 2R\right)  \wspan s&\geq (3+c)\len s-\frac {10rL}{R}
\geq (3+c)\len s-\frac {10L}{R} (\frac {\wspan s}{2L}+4)
\end{align*}
The lemma then follows from the collecting together the two terms involving $\wspan w_1'+\wspan w_2'$, and
then dividing by $1+\frac {7}{R}$.
\end{proof}

The transition to allow some well-matched symbols is done as in
the clean construction and we get the lemma below.
The proof is analogous to that of Lemma~\ref{lem:notbadly} and in particular we
remove the well matched symbols which is shortening $s$ by at most
$4L \cdot \LCS(w_1,w_2)$ and the rest of the proof is essentially identical.
\begin{lemma}
\label{lem:notbadly-dirty}
Suppose $w_1,w_2$ are words over alphabet $[K]$ and $s=(w_1',w_2')$ is a common subsequence between
$\hat{w}_1$ and $\hat{w}_2$. Then
\[
  \wspan s\geq (3+c-\frac{28}{R})\len s-16L\cdot \LCS(w_1,w_2)-\frac {40L}{R}.
\]
\end{lemma}

We are now ready to prove the alphabet reduction claim (Theorem~\ref{thm:alphabet-reduction-dirty})  via concatenation with the dirty construction at the inner level.

\begin{proof}[Proof of Theorem~\ref{thm:alphabet-reduction-dirty} (for binary case)]
All that remains to be done is to pick parameters suitably. 
We set $R$ to the smallest number greater than
$\frac {56}{\gamma}$ such that it can be written on the form 
$(1+c)t$ for and integer $t$ and $c \in [\sqrt 2 -1-\frac {\gamma}{4},
\sqrt 2 -1]$ and we use this value of $c$. It is not difficult to see
that this is possible with $R \in O(\frac 1{\gamma})$. 
Define $T = 2L$ (recall that $L=R^{2K+1}$) and $\tau \colon [K] \to [2]^T$ as $\tau(l) = g_l$ (as defined in \eqref{eq:K-to-k-dirty}), and let $C_2 \subseteq [2]^N$, where $N=2nL$, be the code obtained as in the statement of Theorem~\ref{thm:alphabet-reduction-dirty}. 

By Lemma~\ref{lem:notbadly-dirty}, we can conclude that any common subsequence $s$ of two distinct codewords of $C_2$ satisfies
\[ \wspan s \ge (2+\sqrt 2 -\gamma) \len s - 8\gamma N - \gamma N \ . \]
Since $\len s \le N$, the right hand side is at least $(2+\sqrt 2) \len s - 10 \gamma N$, as claimed in \eqref{eq:dirty-alphabet-reduction}. 
\end{proof}

\begin{remark}
\label{rem:dirty}
For the level of dirt discussed here, i.e., $c \leq \sqrt 2 -1$, the analysis
is optimal for the same reason as the clean one is optimal, as the analysis shows
that the dirt is dropped in forming the subsequence.  Indeed, in the clean construction the efficient 
LCS of length $t$ spans $2t$ symbols in the high frequency string and $t$
symbols in the low frequency string.  Introducing dirt increases the second
number to $t(1+c)$ for a total span of $(3+c)t$.   If the value of $c$ is larger, then the efficient
LCS is obtained by using all symbols, including the dirt, in the low frequency (high amplitude) string.  In
the high frequency string it spans around 
\[
\frac 12 ((1+c)+(1+c)/c) t
\]
symbols (half of the time we are taking the most common symbol, moving at
speed $(1+c)$ and half the time the other symbol moving at speed $(1+c)/c$).
Thus in this case the total span is $\approx  t + (1+c)(1+1/c) t/2 = (2+(c+1/c)/2)t$ and the threshold of $(\sqrt{2}-1)$ for
$c$ was chosen to maximize $\min (3+c,(2+(c+1/c)/2))$.
\end{remark}

\subsubsection{Dirty construction, general case}

Let us give the highlights of the general construction for alphabet size $k$.
In this case we define ``$M$ dirty ones at
frequency $a$'' to be the string
\[
(1^a 2^{ca} 3^{ca} \ldots k^{ca})^{M/(1+(k-1)c)a},
\]
where we assume that $c$ is positive number bounded from above by
$(\sqrt k-1)/(k-1)$.  We denote this string by
$1^k_{M,a}$ and we have analogous dirty strings of 
other symbols.

 The extension of Lemma~\ref{lemma:inftyw1} is as follows.
\begin{lemma}\label{lemma:inftyw1k}
For $j \in [k]$, let $w_1$ be a string of the form $j^k_{\infty,a}$
and let $s$ be a subsequence of 
$w_2=(1^k_{b_1,b_2}2^k_{b_1,b_2}3^k_{b_1,b_2}\ldots k^k_{b_1,b_2})$, then,
\[
  \wspan_{w_1}{s} +kb_1 \geq (k+1+(k-1)c)\len s-\frac {k^2ab_1}{b_2}.
\]
\end{lemma} 

The proof of this lemma follows along the lines of
Lemma~\ref{lemma:inftyw1} with some obvious modifications.  If
we let $S$ be the number of occurrences of $j$'s in $s$ and $\tilde{S}$ the total 
number of other symbols we get a lower bound for the span of the
form
$$
(1+(k-1)c) S + (1+(k-1)c)\tilde{S}/c- \frac {k^2ab_1}{b_2}.
$$
By the upper bound on $c$ we have
$$(1+(k-1)c)/c \geq k+1+(k-1)c,$$
and we can again focus on $\tilde{S}=0$ where again $S \leq b_1$ establishes
the lemma.  The lemma establishes that the span of subsequences of
coding of unequal symbols is large, and adopting the rest of the 
proof to establish 
Theorem~\ref{thm:alphabet-reduction-dirty} for general $k$ is
straightforward and we omit the details.

\section{Existence and construction of good deletion codes}

In this section, we will plug in good ``outer'' deletion codes over large alphabets into Theorem~\ref{thm:alphabet-reduction-dirty} to derive codes over alphabet $[k]$ that correct a fraction $\approx 1-\frac {2}{k+\sqrt k}$ of deletions.

\subsection{Existential claims}

We start with ``outer'' codes over large alphabets guaranteed to exist
by the probabilistic method.  We use $h(\cdot)$ to denote the binary
entropy function. A similar statement to the random coding argument
below appears in \cite{GW-random15}, but we include the short proof
for completeness.

\begin{lemma}\label{lem:rand}
Suppose $\gamma, r >0$ and integer $K \ge 2$ satisfy
\[
  2r \log K +2 h(\gamma)-\gamma \log K<0.
\]
Then, for all large $n$, there exists a code with $K^{r n}$ codewords in
$[K]^n$ such that $\LCS(w,w') < \gamma n$ for all distinct $w,w'$ in
the code.
\end{lemma}
\begin{proof}
Let $w_1,\dotsc,w_{K^{rn}}$ be a sequence of words sampled from $[K]^n$ independently at random \emph{without replacement}.
For any $i<j$ the joint distribution of $(w_i,w_j)$ is same as of two words independently sampled from $[K]^n$
conditioned on them being distinct. Hence, by the union bound we have
\[
  Pr[\LCS(w_i,w_j)>\gamma n]\leq \binom{n}{\gamma n}^2 K^{-\gamma n}.
\]
By the second application of the union bound we thus have
\[
  \Pr[\exists w,w'\in \mathcal{C}_0,\ \LCS(w,w') \ge \gamma n]\leq K^{2rn} \binom{n}{\gamma n}^2 K^{-\gamma n}= 2^{n(2r \log K +2 h(\gamma)-\gamma \log K)+o(n)}<1,
\]
for sufficiently large $n$.
As this probability is less than $1$, there is a choice of $w_1,\dotsc,w_{Mn}$ such that pairwise $\LCS$ is less than $\gamma n$.
\end{proof}

Using the above existential bound in Theorem~\ref{thm:alphabet-reduction-dirty}, we now deduce the following.

\begin{theorem}[Existence of deletion codes]\label{thm:main-existential}
Fix an integer $k\geq 2$. Then for every real number $\veps>0$, there is $\tilde{r}=(\veps/k)^{O(\veps^{-3})}$ such that for infinitely many $N$ there is a code $C \subseteq [k]^N$  of rate at least $\tilde{r}$ and $\LCS(C) < \left(\frac{2}{k+\sqrt k} + \veps\right) N$.
\end{theorem}
\begin{proof}
We first apply Lemma~\ref{lem:rand} with $\gamma = \eps/4$ and $r = \gamma/6 = \eps/24$ to get a code $C_1 \subseteq [K]^n$ for $K \le O(1/\eps^3)$ with $\LCS(C_1) \le \eps n/4$ and $|C_1| \ge K^{rn}$. Now applying Theorem~\ref{thm:alphabet-reduction-dirty} to $C_1$ yields a code $C_2 \subseteq [k]^N$ with $\LCS(C_2) \le \left(\frac{2}{k+\sqrt k} + \veps\right) N$. The rate of $C_2$ is at least $r/T \ge  (\veps/k)^{O(\veps^{-3})}$ since $T \le (k/\eps)^{O(K)}$.
\end{proof}

\begin{remark}
The exponent $O(1/\eps^3)$ in the rate can be improved to $O(1/\eps^a)$ for any $a > 2$. We made the concrete choice $a=3$ for notational convenience.
\end{remark}

\subsection{Efficient deterministic construction}\label{subsec:efficient}

Theorem~\ref{thm:main-existential} already shows the {\em existence}
of positive rate codes over the alphabet $[k]$ which are capable of
correcting a deletion fraction approaching $1-\frac{2}{k+\sqrt k}$, giving our
main combinatorial result. We now turn to explicit constructions of
such codes. Given Theorem~\ref{thm:alphabet-reduction-dirty}, all that we
need is an explicit code family capable of correcting a deletion
fraction approaching $1$ over constant-sized alphabets, which is
guaranteed by the following theorem.

\begin{lemma}[\cite{GW-random15}, Thm 3.4]
\label{lem:GW-codes}
For every $\gamma > 0$ there exists an integer $K \le O(1/\gamma^5)$
such that for infinitely many block lengths $n$, one can construct a
code $C \subseteq [K]^n$ of rate $\Omega(\gamma^3)$ and $\LCS(C) \le
\gamma n$ in time $n (\log n)^{\mathrm{poly}(1/\gamma)}$. Further, the
code $C$ can be efficiently encoded and decoded from a fraction
$(1-\gamma)$ of deletions in $n \cdot (\log
n)^{\mathrm{poly}(1/\gamma)}$ time.
\end{lemma}

\begin{remark}
The linear dependence on $n$ in the decoding time can be deduced using
fast ($n \cdot \mathrm{poly}(\log n)$ time) unique decoding algorithms for
Reed--Solomon codes. The bounds stated in \cite{GW-random15} are
$n^{O(1)} (\log n)^{\mathrm{poly}(1/\gamma)}$ time.
\end{remark}

Using the efficiently constructible codes of Lemma~\ref{lem:GW-codes}
in place of random codes as outer codes, we can
get the constructive analog of Theorem~\ref{thm:main-existential} with
a similar proof. We also record the statement concerning the span of
common subsequences of distinct codewords of our code (which is guaranteed by Theorem~\ref{thm:alphabet-reduction-dirty}), as we will make
use of this in the next section on efficiently decodable deletion
codes.

\begin{theorem}[Constructive deletion codes]
\label{thm:deletion-codes-constructive}
Fix an integer $k\geq 2$. Then for every real number $\veps>0$, there
is $\tilde{r}=(\veps/k)^{O(\veps^{-3})}$ such that for infinitely many
$N$, we can construct a code $C \subseteq [k]^N$ in time $O(N (\log
N)^{\mathrm{poly}(1/\eps)})$ such that (i) $C$ has rate at least
$\tilde{r}$ and (ii) $\LCS(C) < \left(\frac{2}{k+\sqrt k} + \veps\right) N$;
in fact if $s$ is a common subsequence of two distinct codewords $c,\tilde{c}
\in C$, then $\wspan s \ge (k+\sqrt k)\len s - \eps k N$.

\end{theorem}

\section{Deletion codes with efficient decoding algorithms}\label{sec:decoding}

We have already shown how to efficiently construct codes over alphabet
$[k]$ that are combinatorially capable of correcting a deletion
fraction approaching $1 -\frac{2}{k+\sqrt k}$. However, it is not so clear
how to efficiently recover the codes in
Theorem~\ref{thm:deletion-codes-constructive} from deletions. To this
end, we now give an alternate explicit construction by concatenating codes with large distance for the {\em Hamming metric} with good $k$-ary deletion codes as constructed in the previous section.
As a side benefit, the construction time will be improved as we will need the codes from Theorem~\ref{thm:deletion-codes-constructive} for exponentially smaller block lengths.


\subsection{Concatenating Hamming metric codes with deletion codes}
We state our concatenation result abstractly below, and then instantiate with appropriate codes later for explicit constructions. Recall that the relative distance (in Hamming metric) of a code $C$ of block length $n$ equals the minimum value of $\Delta(c,\tilde{c})/n$ over all distinct codewords $c,\tilde{c} \in C$, where $\Delta(x,y)$ denotes the Hamming distance between two words of the same length.
\begin{lemma}
\label{lem:hamming+deletion}
Let $\eta,\theta \in (0,1]$.
Let $C_{\rm out} \subseteq [Q]^n$ be code of relative distance in Hamming metric at least $(1-\eta)$. Let $C_{\rm in} \subseteq [k]^m$ be a code with $nQ$ codewords, one for each $(i,\alpha) \in [n] \times [Q]$, such that for any two distinct codewords $c_1,c_2  \in C_{\rm in}$ and a common subsequence $s$ of $c_1,c_2$, we have
$\wspan s \ge (k+1) \len s - \theta k m$. Consider the code $C_{\rm concat} \subseteq [k]^{N}$ for $N=nm$ obtained as follows\footnote{Note that this is a concatenation of a ``position-indexed version" of $C_{\rm out}$ with $C_{\rm in}$.}: There will be a codeword of $C_{\rm concat}$ for each codeword $c$ of $C_{\rm out}$, obtained by replacing its $i$'th symbol $c_i$ by the codeword of $C_{\rm in}$ corresponding to $(i,c_i)$.
Then we have
\[ \LCS(C) \le \left( \frac{2}{k+\sqrt k} + 2\theta + \eta \right) N  \ . \]
\end{lemma}
\begin{proof}
This proof is similar to, but simpler than the proofs of Lemmas~\ref{lem:badly} and~\ref{lem:notbadly}. It is simpler because in the present
situation a codeword of $C_{\rm in}$ occurs at most once inside a codeword of $C_{\rm concat}$.

Let $c,\tilde{c}$ be two distinct codewords of $C_{\rm concat}$ and let $\sigma$ be a common subsequence of $c,\tilde{c}$. Recall that each codeword of $C_{\rm concat}$ can be viewed as a sequence of $n$ (inner) blocks belonging to $[k]^m$, with the $i$'th block encoding (as per $C_{\rm in}$) the $i$'th symbol of the outer codeword. Let us break $\sigma$ into parts based on which of the $n$ blocks in $c,\tilde{c}$ its common symbols come from in some canonical (say greedy) way of forming the subsequence $\sigma$ from $c,\tilde{c}$).  Let $\sigma_{i,j}$ denote the portion of $\sigma$ formed by using symbols from the $i$'th block of $c$ and the $j$'th block of $\tilde{c}$. Let $E$ be the set of pairs $(i,j)$ for which $\sigma_{i,j}$ is not the empty word. If we were to draw words $c$ and $\tilde{c}$ as horizontal lines parallel to each other with the $n$ blocks marked as vertically aligned points on the lines, and draw the pairs in $E$ as edges between corresponding points, then they would be non-crossing.
Therefore, $|E| \le 2n$.
Also, by the construction, the only portions $\sigma_{i,j}$ that are formed out of the same codeword of $C_{\rm in}$ are those with $i=j$ and $c_i=\tilde{c}_i$. Thus there are at most $\eta n$ such portions, by the assumed relative distance of $C_{\rm out}$. Combining all this, we have
\begin{align*}
\wspan \sigma & \ge \sum_{(i,j) \in E} \wspan \sigma_{i,j} \\
& \ge \left( \sum_{(i,j) \in E} \Bigl( (k+\sqrt k) \len \sigma_{i,j} - \theta k m  \Bigr) \right) - (k+\sqrt k) (\eta n) m \\
& \ge (k+\sqrt k) \len \sigma - 2\theta k nm - (k+\sqrt k) \eta nm \ .
\end{align*}
Since $\wspan \sigma \le 2N$, we have
$\len \sigma < \left( \frac{2}{k+\sqrt k} + 2 \theta + \eta \right) N$, as desired.
\end{proof}

\smallskip \noindent {\bf The construction.}
We now instantiate the above by concatenating Reed--Solomon codes with the codes from
Theorem~\ref{thm:deletion-codes-constructive}. Fix the desired alphabet size $k \ge 2$ and $\gamma > 0$.

Let $\F_q$ be a large finite field,
an integer $\ell = \lceil \frac{\gamma q}{2} \rceil$.
Let $C_{\rm out}$ be the  Reed--Solomon
encoding code of block length $n=q$ that maps degree
$< \ell$ polynomials $f \in \F_q[X]$ to their evaluations at all points in $\F_q$. Note that its relative distance is $(q-\ell+1)/q \ge 1- \gamma/2$.

 Let $C_{\rm in}$ be a $k$-ary code with at least $q^2$ codewords constructed in Theorem~\ref{thm:deletion-codes-constructive} for $\eps = \gamma/4$. By the promised rate of that construction, the block length of $C_{\rm in}$ can be taken to be $m \le (k/\gamma)^{O(\gamma^{-3})} \cdot \log q$.
Our final construction will apply Lemma~\ref{lem:hamming+deletion} to $C_{\rm out}$ and $C_{\rm in}$ with parameters $\eta = \gamma/2$ and $\theta = \gamma/4$, to get a code $C_{\rm concat} \subseteq [k]^N$ for $N=qm$ with $\LCS(C_{\rm concat}) \le \left( \frac{2}{k+\sqrt k} + \gamma\right) N$.

Let us now estimate the construction time. As a function of $N$, $m \le O_{k,\gamma}(\log N)$, and therefore the construction time for $C_{\rm in}$ becomes $O_{k,\gamma}(\log N (\log \log N)^{\mathrm{poly(1/\eps)}})$. Together with the $q (\log q)^2$ time to construct a representation of $\F_q$ and the Reed--Solomon code, we get an overall construction time of $O(N \log^2 N)$ for large enough $N$.  We record this in the following statement.

\begin{theorem}[Reed--Solomon + inner deletion codes with better construction time]
\label{thm:RS-concat}
Fix an integer $k\geq 2$. Then for every real number $\gamma >0$, there is $r(k,\gamma)=(\gamma/k)^{O(\gamma^{-3})}$ such that for infinitely many and sufficiently large $N$, we can construct a code $C \subseteq [k]^N$ in time $O(N \log^2 N)$ such that 
\begin{enumerate}
\item[(i)] $C$ has rate at least $r(k,\gamma)$, and 
\item[(ii)] $\LCS(C) < \left(\frac{2}{k+\sqrt k} + \gamma \right) N$.
\end{enumerate}
\end{theorem}

\subsection{Deletion correction algorithm}
We now describe an efficient decoding procedure for the codes from Theorem~\ref{thm:RS-concat}. The procedure will succeed as long as the fraction
of deletions is only slightly smaller than $1-\frac{2}{k+\sqrt k}$. We describe the basic idea before giving the formal statement and proof. If we are given a subsequence $s$ of length $\bigl( \frac{2}{k+\sqrt k} + \delta\bigr)N$ of some codeword, then by a simple counting argument, there must be at least $\delta q/2$ inner blocks (corresponding to the inner encodings of the $q$ indexed Reed--Solomon symbols) in which $s$ contains at least  $\bigl( \frac{2}{k+\sqrt k} +\frac{\delta}{2} \bigr)m$  symbols from the corresponding inner codeword. So we can decode the corresponding Reed--Solomon symbol (by brute-force) if we knew the boundaries of this block. Since we do not know this, the idea is to try decoding all contiguous chunks of size $\bigl(\frac{2}{k+\sqrt k} + \frac{\delta}{4}\bigr)m$ in $s$ with sufficient granularity (for example, subsequences beginning at locations which are multiples of $\delta m/4$).

This might result in the decoding of several spurious symbols, but there will be enough correct symbols to {\em list decode} the Reed--Solomon code and produce a short list that includes the correct message. By the combinatorial guarantee on the LCS value of the concatenated code from Theorem~\ref{thm:RS-concat}, only the correct message will have an encoding containing $s$ as a subsequence. Therefore, we can prune the list and identify the correct message by re-encoding each candidate message and checking which one has $s$ as a subsequence.
The list decoding step is similar to the one used in \cite{GW-random15} for list decoding binary codes from a fraction of deletions approaching $1/2$. Since we have the combinatorial guarantee that the code can correct a deletion fraction $\approx 1- \frac{2}{k+\sqrt k}$, a list decoding algorithm up to this radius is also automatically a unique decoding algorithm.

\begin{theorem}[Explicit and efficiently decodable deletion codes]
\label{thm:final-efficient-decoding}
The concatenated code $C \subseteq [k]^N$ constructed in Theorem~\ref{thm:RS-concat} can be efficiently decoded from a fraction $\bigl( 1 - \frac{2}{k+\sqrt k} - O(\gamma^{1/3}) \bigr)$ of worst-case deletions in $N^3 (\log N)^{O(1)}$ time, for large enough $N$.
\end{theorem}
\begin{proof}
With hindsight, let $\delta=3\gamma^{1/3}$.
Suppose we are given a subsequence $s$ of an {\em unknown} codeword $c \in C$ (encoding the unknown polynomial $f$ of degree $< \ell$), where $\len s \ge \bigl( \frac{2}{k+\sqrt k} + \delta \bigr) N$.
We claim that the following decoding algorithm recovers $c$.
\begin{enumerate}
\itemsep=0ex
\item $\mathcal{T} \leftarrow \emptyset$.
\item ~[Inner decodings] For each integer $j$, $0 \le j \le \frac{\len s}{(\delta m)/4}$, do the following:
\begin{enumerate}
\itemsep=0ex
\item Let $\sigma_j$ be the contiguous subsequence of $s$ of length $\bigl( \frac{2}{k+\sqrt k} + \frac{\delta}{4} \bigr) m$ starting at position $ j \lfloor \frac{\delta m}{4} \rfloor + 1$.
\item By a brute-force search over $\F_q \times \F_q$, find the unique pair $(\alpha,\beta)$, if any, such that its encoding under $C_{\rm in}$ has $\sigma_j$ as a subsequence, and add $(\alpha,\beta)$ to $\mathcal{T}$. (This pair, if it exists, is unique since $\LCS(C_{\rm in}) < \bigl( \frac{2}{k+\sqrt k} + \frac{\gamma}{4} \bigr) m$, and $\delta \ge \gamma$.)
\end{enumerate}
\item ~[Reed--Solomon list recovery] Find the list, call it $\mathcal{L}$, of all polynomials $p \in \F_q[X]$ of degree $< \ell$ such that
\begin{equation}
\label{eq:rs-ld-cond}
\Big| \{ (\alpha, p(\alpha)) \mid \alpha \in \F_q \} \cap \mathcal{T} \Big| \ge \frac{\delta q}{2} \ .
\end{equation}
\item ~[Pruning] Find the unique polynomial $f\in \mathcal{L}$, if any, such that its encoding under $C$ contains $s$ as a subsequence, and output $f$.
\end{enumerate}

\noindent {\sc Correctness.}  Break the codeword $c \in [k]^{nm}$ of the concatenated code $C$ into
$n$ (inner) blocks, with the $i$'th block $b_i \in [k]^m$ corresponding to the
inner encoding of the $i$'th symbol $(\alpha_i, f(\alpha_i))$ of the
outer Reed--Solomon codeword.   For some fixed canonical way of
forming $s$ out of $c$, denote by $s_i$ the portion of $s$
consisting of the symbols in the $i$'th block $b_i$.
Call an index $i \in [n]$ \emph{good} if
$\len s_i \ge \left( \frac{2}{k+\sqrt k} + \frac{\delta}{2} \right) m$. By a
simple counting argument, there are at least $\delta n/2$ values of $i \in [n]$ that are good.

For each good index $i \in [n]$, one of the inner decodings in Step 2 will attempt to decode a subsequence of $s_i$, and therefore will find the pair $(\alpha_i,f(\alpha_i))$. Since there are at least $\frac{\delta q}{2}$ good indices, the condition \eqref{eq:rs-ld-cond} is met for the correct $f$. Using Sudan's list decoding algorithm for Reed--Solomon codes~\cite{sudan}, one can find the list of all degree $\le \ell$ polynomials $p \in \F_q[X]$ such that $(\alpha,p(\alpha)) \in \mathcal{T}$ for more than $\sqrt{2 \ell |\mathcal{T}|}$ field elements $\alpha \in \F_q$.  Further, this list will have at most $\sqrt{2 |\mathcal{T}|/\ell}$ polynomials.

Since $|\mathcal{T}| \le 4q/\delta$, if we pick $\delta$ so that $\frac{\delta q}{2} > \sqrt{ 8 \ell q/\delta}$, the decoding will succeed. Recalling that $\ell = \lceil \frac{\gamma q}{2} \rceil$, this condition is met for our choice of $\delta$.\smallskip

\noindent {\sc Runtime.} The number of inner decodings performed is
$O(q/\delta) = O(N)$, and each inner decoding takes $q^2 (\log
q)^{O(1)} \le N^2 (\log N)^{O(1)})$ time. The set $\mathcal{T}$ has
size at most $O(q/\delta) \le O(N)$ for $N$ large enough. The
Reed--Solomon list decoding algorithm on $|\mathcal{T}|$ many points
can be performed in $O(N^2)$ field operations, see for instance
\cite{roth-ruckenstein}. So the overall running time of the decoder is
at most $N^3 \cdot \mathrm{poly}(\log N)$.
\end{proof}

\begin{remark}
The cubic runtime in the above construction arose because of the brute-force implementation of the inner decodings.
One can recursively use the above concatenated codes themselves as the inner codes, in place of the codes from Theorem~\ref{thm:deletion-codes-constructive}.  Each of the inner decodings can now be performed in $\mathrm{poly}(\log q)$ time, for a total time of $N \cdot \mathrm{poly}(\log N)$ for Step 2. By using near-linear time implementations of Reed--Solomon list decoding~\cite{alekhnovich05}, one can also perform Step 3 in $q \cdot \mathrm{poly}(\log q)$ time. Thus one can improve the decoding complexity to $N \cdot \mathrm{poly}(\log N)$.
\end{remark}

\section{Concluding remarks}

The obvious question left open by this work is to determine the exact 
value of $p^*(k)$, the (supremum of the) largest fraction of deletions one can correct over alphabet size $k$ with positive rate. Even in the binary case we do not dare to have a strong 
opinion whether the value is $\frac 12$, $\sqrt 2-1$ or some intermediate
value, but let us close with a few comments.

When comparing the encodings of
two different symbols in our inner code, one codeword looks locally like
$1^A 2^{cA}$ (or the other way around) 
where the other codeword has long
stretches (of length $\gg A$) of the same symbol (which are equally often 1's and 2's).  
It is tempting to introduce one more level of granularity, let us call
it ``micro particles'' in these long stretches, in the form of
sequences of the form $j^B$ for $j \in \{ 1,2\}$ and $B$ smaller
than $A$.  We were unable to use this to improve the bounds of
the contruction.  It seems like only the shortest period in each of the
two codewords matter but we do not have a formal statement to support
this feeling.

There are two reasons for subsequences having big spans in our
construction.  The first reason is that the frequencies
are different (this is the main mechanism in the clean construction
and hence in \cite{BG-soda16}) and the second is the impurities in the form of dirt.
The span is large because we discard half of the high frequency string
and all of the dirt.  If the span is to approach 4 times the length of the
subsequences, we need the fraction of dirt to approach half the length
of the string but this seems hard to combine with the intuition of
being ``dirt,'' which should be in minority.  We suspect that
some new mechanism is needed to prove that $p^*(2)= \frac 12$ 
if this is indeed the true answer.

\bibliographystyle{plain}
\bibliography{deletioncodes}

\end{document}